\documentclass[prd,aps,showpacs,nofootinbib,preprint,eqsecnum]{revtex4}
\usepackage{graphicx}
\usepackage{epsfig}
\usepackage{bm}
\usepackage{amsfonts}
\usepackage{graphicx,color,amsmath,amsxtra}
\usepackage{epsf}
\usepackage{amssymb}
\usepackage{enumerate}
\usepackage{hhline}
\usepackage{array}
\usepackage{tabularx}

\newcommand{\be}{\begin{equation}}
\newcommand{\ee}{\end{equation}}
\newcommand{\bea}{\begin{eqnarray}}
\newcommand{\eea}{\end{eqnarray}}
\newcommand{\beaa}{\begin{eqnarray*}}
\newcommand{\eeaa}{\end{eqnarray*}}

\newcommand{\Eqn}[1]{&\hspace{-0.2em}#1\hspace{-0.2em}&}

\def\be{\begin{equation}}
\def\ee{\end{equation}}
\def\bea{\begin{eqnarray}}
\def\eea{\end{eqnarray}}

\begin{document}

\title{Kaluza-Klein reduction and Bergmann-Wagoner bi-scalar general action of scalar-tensor gravity} 

\author{Kazuharu Bamba$^{1, 2, 3}$, 
Davood Momeni$^{4}$
and 
Ratbay Myrzakulov$^{4}${}}

\affiliation{
$^1$Division of Human Support System, Faculty of Symbiotic Systems Science, Fukushima University, Fukushima 960-1296, Japan
\\
$^2$Leading Graduate School Promotion Center, 
Ochanomizu University, Tokyo 112-8610, Japan
\\
$^3$Department of Physics, Graduate School of Humanities and Sciences, 
Ochanomizu University, Tokyo 112-8610, Japan
\\
$^4$Eurasian International Center for Theoretical Physics, Eurasian National University, Astana 010008, Kazakhstan
}


\begin{abstract}
We examine the Kaluza-Klein (KK) dimensional reduction from higher-dimensional space-time and the properties of the resultant Bergmann-Wagoner general action of scalar-tensor theories. 
With the analysis of the perturbations, we also investigate the stability of the anti-de Sitter (AdS) space-time in the $D\in\mathcal{N}$-dimensional Einstein gravity with the negative cosmological constant. Furthermore, we derive the conditions for the dimensional reduction to successfully be executed and present the KK compactification mechanism.
\end{abstract}

\preprint{OCHA-PP-321, FU-PCG-4}

\maketitle

\section{Introduction} 

According to recent cosmological observations, 
not only inflation in the early universe but also the late-time 
cosmic acceleration (the so-called dark energy problem) has been supported. 
As a possible approach to realize these cosmic accelerated expansions, 
recently, scalar-tensor theories including the (Jordan-Fierz-)Brans-Dicke (BD) theory~\cite{B-D} has widely been studied in the literature. 
This is because modified gravity theories such as $F(R)$ gravity 
can be rewritten to a kind of the BD type theory , called as Bergmann and Wagoner theory
(for reviews on issues of dark energy and modified gravity, 
see, for instance,~\cite{R-MG, Bamba:2012cp}). 
It is known that scalar-tensor theories can be 
constructed as four-dimensional effective theories 
through the Kaluza-Klein (KK) dimensional reduction from higher-dimensional space-time theories (for detailed reviews on the KK mechanism of the dimensional reduction, see, e.g.,~\cite{K-K, Blagojevic:2002du, I-M, 
Vladimirov:1999vh, B-R}). 
It is remarkable to mention that the modification of the BD theory 
in an arbitrary dimensional space-time has been studied 
in Ref.~\cite{Rasouli:2014dxa}, 
and that the application of the KK dimensional reduction from the 
five-dimensional space-time to the so-called Eddington-Born-Infeld action 
has been executed in Ref.~\cite{Fernandes:2014bka}. 

Moreover, in the framework of the anti-de Sitter (AdS)/the conformal field theory (CFT) correspondence~\cite{AdS-CFT}, with the KK dimensional reduction, 
the relation between the solutions of general relativity plus 
a negative cosmological constant and those in the Minkowski space-time 
has been explored~\cite{C-C-G-S}. 
Furthermore, very recently, by developing the above considerations, 
the connection between the solutions of general relativity 
with a positive cosmological constant and those of that with 
a negative one in Ref.~\cite{DiDato:2014kca}. 

In this paper, we explore the procedure of the KK dimensional reduction from higher-dimensional space-time. In particular, we investigate the natures of the scalar-tensor theories obtained through the KK compactification\footnote{Aa a phenomenological application of the KK compactification mechanism, the first KK mode of neutrino has been considered to be the cold dark matter in Ref.~\cite{Nishio:2014sva}.}. Concretely, by starting with the Einstein-Hilbert action in the $D$-dimensional space-time ($D \geq 5$) and adopting the compactification of the coordinate, we find the  Bergmann-Wagoner formulation of the BD theory in $(D-1)$-dimension \cite{BW}. 
Indeed, we find a class of general action of scalar-tensor gravity with two scalar fields which is at most quadratic in derivatives of the fields.

The resultant theory is different from the usual BD one, because instead of the usual BD theory, we also have an auxiliary scalar field, $\Psi$, 
which is non-minimally coupled to the BD scalar field\footnote{In 
Ref.~\cite{Bamba:2006mh}, cosmology in the theory with two scalar fields non-minimally coupled to the Ricci scalar has been studied in detail.}. 
In the literature it was called as Bergmann-Wagoner formulation of scalar-tensor gravity with one scalar field. The general action of Bergmann-Wagoner action is simply written in the Jordan frame\footnote{We mention here that according to Ref.~\cite{Y. M. Cho}, 
the quantum fluctuations violate the weak equivalence principle in 
the BD theory. This means that the Jordan frame can no longer be interpreted as physical.}:
\begin{eqnarray}
S=\frac{1}{2\kappa_{D}^2}\int{d^Dx\sqrt{-g}\Big[\phi R-\frac{\omega(\phi)}{\phi}g^{\mu\nu}\partial_{\mu}\phi\partial_{\nu}\phi-U(\phi)\Big]}+S_\mathrm{M}\Big[\Psi,g_{\mu\nu}\Big],
\end{eqnarray}
where $\omega$ and $U$ are arbitrary functions of the scalar field $\phi$, and $S_\mathrm{M}$ stands out for its matter sector. When $\omega=\omega_\mathrm{BD}$ is constant, $U(\phi)=0$, the model is called as BD gravity \cite{Fierz}. The alternative action which we found is totally different from BD. We showed that by KK reduction, the Bergmann-Wagoner action is reduced a bi-scalar model $\phi,\psi$. The second scalar field $\psi$ is non-minimally coupled to the first one $\phi$ and gravity $g_{\mu\nu}$. By taking into account the Bergmann-Wagoner formulation of BD, it is adequate to call it as the Bergmann-Wagoner bi-scalar general action of scalar-tensor gravity\footnote{It is remarkable to explicitly state that 
the reduction process from higher-dimensional theories 
performed in Ref.~\cite{DiDato:2014kca} corresponds to 
a specific class of our investigations.} 

The methodology in this work seems to be similar to the KK compactification, but there exists the difference between our approach and the KK procedure 
in terms of the technique and result. 
Our model of lower-dimensional gravity includes two types of the auxiliary fields. A non-minimal coupling between the fields appears in a natural way and in a systematic form. Here, such a non-minimal interaction is not introduced ad hoc. So, consequently the Bergmann-Wagoner action under KK reduction reduces to the bi-scalar model.
Another motivation of the present work is to extend the KK compactification to more than one compactified direction. As a natural extension, if we start with the $D\geq5$ -dimensional  metric and we compactified $D-4$-coordinates $y^i \sim y^i+2\pi R^i$ with $1 \leq i \leq D-4$, the reduction of the Einstein-Hilbert action to the lower-dimensional space-time is a type of the Bergmann-Wagoner theory with a non-minimally coupling in $(D-1)$-dimensional space-time. We here emphasize that our dimensional reduction scheme is different from the original reductions of KK, Scherk-Schwarz (I, II). 
One more reason for the study of such dimensional reductions of gravity in higher-dimensional space-time is that 
all of the most reliable unified models of the fundamental four forces in the nature such as superstring ($D=10$) and M-theory ($D=11$) live in the higher-dimensional $D\geq5$ space-time. 
We use units of $k_{\mathrm{B}} = c = \hbar = 1$ and denote the 
gravitational constant, $G$, by $\kappa^2\equiv 8 \pi G$, so that
$G=1/M_{\mathrm{Pl}}^2$ with $M_{\mathrm{Pl}} =1.2 \times 10^{19}$ GeV the Planck mass. 

The organization of the paper is the followings. 
In Sec.\ II, we explain the formulation of the KK reduction mechanism from and 
study the natures of the resultant theories. The field equations are also derived. 
In Sec.\ III, we explore the conditions for the procedure of the dimensional reduction to be performed successfully and mention the KK compactification 
mechanism. 
In Sec.\ IV, we investigate the stability of the AdS solution through the analysis of the perturbations around the background AdS solution. 
In Sec.\ V, conclusions are described.  

\section{Reduction formalism}

We first present the formalism of the KK dimensional reduction. 

\subsection{Formulation} 

We use the following notations for indices. 
The Greek indices run as $\mu,\, \nu = 0, \dots, D$ and the Latin indices do like $i,\, j, \dots = 1, \dots, D$, where $n$ is a natural number. 
The coordinate frame is defined by $x^{\mu} \equiv \{x^0,x^i\}$. Metric of the space-time is defined by the following representation:
\begin{equation}
g=g_{\mu\nu}dx^{\mu}\otimes dx^{\nu}.
\end{equation}

Suppose that the metric $g_{\mu\nu}$ of the space-time is static with respect to a coordinate, we call it as $x^0=t$. 
In general, it does not mean the physical time, but it can be angle of only a coordinate without any meaning of length. However, since we usually use the terminology ``static'' for the time-independent and irrotational sources of matters, we use the same notation as time. 
For the static metric in this sense, it satisfies the two conditions 
\begin{equation}
\frac{\partial g_{\mu\nu}}{\partial t}=0\,, 
\quad 
g_{it}=0\,.
\end{equation} 
Here, owing to the nature of being non-stationary, 
there is no gauge field $A_{\mu}=A_{t}$ like in the case of the KK reduction, and therefore the model has no $U(1)$ gauge field. 
In the language of the ($1 + 3$) time-space decomposition~\cite{LL}, 
we have only gravitoelectric field. In the original KK reduction, 
we have gravitoelectromagnetic fields.
\par

We adopt the static metric to the parameterization 
\begin{equation}
V_D:\ \ g=g_{\mu\nu}dx^{\mu}\otimes dx^{\nu}=-e^{2\gamma}dt\otimes dt
+e^{2\sigma}h_{ij}dx^i\otimes dx^j \,.
\label{g}
\end{equation}
We assume that $h_{ij}$ is also static and metric functions are 
time-independent. In this case, It defines the space-time
\begin{equation}
V_{D-1}:\ \ dl^2=h_{ij}dx^i\otimes dx^j \,.
\label{h}
\end{equation}
Clearly, we see that 
\begin{equation}
V_{D-1}\subset V_{D} \,.
\end{equation}
Our aim here is to reduce the Einstein-Hilbert action in terms of $D$-dimensional metric to a lower dimensional action. The action in the $D$-dimensional 
space-time is described as 
\begin{equation}
S_{D}=\int \sqrt{-g} d^D x \frac{R_{D}}{2\kappa_D^2}\,, 
\label{Sn}
\end{equation}
where $R_{D}$ is the Ricci scalar in the $D$-dimensional space-time 
and $\kappa_D^2 \equiv 8 \pi G_D$ with $G_D$ the gravitational constant 
in the $D$-dimensional space-time. In what follows, the subscription 
$D$ ($(D-1)$) denotes the quantities in the $D$ ($(D-1)$)-dimensional 
space-time. 
With the metric in Eq.~(\ref{g}), 
the Ricci scalar $R (V_D)=R_D$ 
is given by 
\begin{eqnarray}
R_{D} \Eqn{=} 
e^{-2\sigma}\left[ 
R_{D-1}+2(D-2)\bigtriangleup \sigma+2\bigtriangleup\gamma+2(D-3)
\partial_i\sigma\partial^i\gamma \right.
\nonumber \\ 
&&
\left.
\hspace{10mm}
+(D-2)(D-3)\partial_i\sigma\partial^i\sigma+
2\partial_i\gamma\partial^i \gamma \right]\,.
\label{Rn}
\end{eqnarray}
This is a recursion relation between $R_D$ and $R_{D-1}$. 
For a given set of the metric functions, it is not easy to find $R_{D-1}$. The only simple solvable case is happen when $\sigma=0$. 
%
%
{}From this recursion relation, we obtain 
the following formal solution for $R_D$: 
\begin{eqnarray}
R_D \Eqn{=} 
e^{-2\Sigma_{l=0}^{k-1}\sigma_{D-l}}R_{D-k}+f_{D,D-1}
+e^{-2\sigma_D}f_{D-1,D-2} \nonumber \\ 
&& 
{}+e^{-2(\sigma_D+\sigma_{D-1})}f_{D-2,D-3}
+e^{-2(\sigma_D+\sigma_{D-1}+\sigma_{D-2})}f_{D-3,D-4}
+ \dots  \nonumber \\ 
&& 
{}+e^{-2(\sigma_D+\sigma_{D-1}+...+\sigma_{D-k/2})}
f_{D-(k-1),D-k}+ \dots \,. 
\end{eqnarray}
We can truncate it at $k=D-4>0$. 
However, still this form is too much complicated to work out well for our theory. Thus, we kept it just as an interesting mathematical problem. 
\par
Here, all the derivatives are with respect to 
$V_{D-1}:\ \ h_{ij}dx^i\otimes dx^j$ in Eq.~(\ref{h}), where
\begin{equation}
\partial_i f=\frac{\partial f}{\partial x^i}\,,
\quad 
\bigtriangleup=h^{-1/2}\partial_{i}(h^{ik}h^{1/2}\partial_{k})\,.
\end{equation}
It is necessary to find the determinant of the $V_D$, form which 
we have 
\begin{equation}
\sqrt{-g}=e^{\gamma+(D-1)\sigma}\sqrt{h}\,.
\label{gn}
\end{equation} 

Now, provided that in the same terminology as in the KK reduction, 
$t$ is called as the compactified direction
\begin{equation}
t\sim t+2\pi R \,.
\label{sn-1}
\end{equation}
We define the two new metric functions 
\begin{equation}
\lambda=\frac{D-3}{2}\sigma \,,
\quad
\mu=\gamma+\lambda \,.
\end{equation}
Consequently, the action in Eq.~(\ref{Sn}) is rewritten 
in terms of these functions as 
\begin{equation}
S_{D-1}=\int{\frac{\sqrt{h}e^{\mu+\lambda}} {2\kappa_D^2}d^{D-1}x \left[R_{D-1}
+2\bigtriangleup\mu+\frac{2(D-1)}{D-3}\bigtriangleup\lambda+\frac{2(D-1)}{D-3}\partial_i\lambda\partial^i\lambda
+2\partial_i\mu\partial^i\mu\right]} \,.
\end{equation}
This is the $(D-1)$-dimensional action with two scalar degrees of freedom. 
With the partial integration over the static boundary $\partial M$, 
the action is simplified as 
\begin{equation}
\int{\sqrt{h}e^{\mu+\lambda}\bigtriangleup\mu  d^{D-1}x}=\int_{\partial M}{\sqrt{h}h^{ij}n_i\partial_j\mu e^{\mu+\lambda}d^{D-2}x}
-\int{h^{1/2}h^{ij}\partial_j\mu (\partial_i\mu+\partial_i\lambda)e^{\mu+\lambda}d^{D-1}x}\,, 
\end{equation}
where $n_i$ is a unit vector normal to the hypersurface defined by 
$x^{D-1} \equiv \text{constant}$. 
As a result, we obtain
\begin{eqnarray}
S_{D-1} \Eqn{=} \int{\sqrt{h}e^{\mu+\lambda}d^{D-1}x \left[ \frac{R_{D-1}}{2\kappa_D^2}-\frac{2(D-2)}{\kappa_D^2(D-3)}\partial_i\mu \partial^i\lambda \right]} 
\nonumber \\
&&
{}+\kappa_{D}^{-2}\int_{\partial M}{\sqrt{h}h^{ij}n_{i}e^{\mu+\lambda}d^{D-2}x
\Big(\partial_j\mu+\frac{D-1}{D-3}\partial_j\lambda\Big)}\,, 
\label{eq:2.13}
\end{eqnarray}
%

\subsection{Features of the resultant lower-dimensional theories}

Furthermore, we define the following two new functions $\Phi$ and 
$\Psi$ by $ \mu+\lambda \equiv \ln \Phi$ and $\mu-\lambda= \ln \Psi$. 
By using these two scalar fields, the action $S_{n-1}$ can eventually 
be represented as 
\begin{equation}
S_{D-1}=\int \frac{\sqrt{h}d^{D-1}x}{2\kappa_D^2}\left[ \Phi \left(R_{D-1}-\frac{\zeta}{4} \frac{\Phi_{,i}\Phi^{,i}}{\Phi^2} \right)+\frac{\zeta\Phi}{4} \frac{\Psi_{,i}\Psi^{,i}}{\Psi^2}) \right] \,,
\label{result1}
\end{equation}
where 
\begin{equation}
\zeta=\frac{2(D-2)}{D-3}\,.
\end{equation}
This is a new type of the non-minimally coupled Lagrangian of the Bergmann-Wagoner formulation of scalar-tensor gravity with one scalar field theory . If we define the coupling constant parameter $\omega_\mathrm{BD} = \frac{\zeta}{4}$, then our action (\ref{result1}) is a type of  Bergmann-Wagoner in the following form:
\begin{eqnarray}
S=\frac{1}{2\kappa_{D}^2}\int{d^{D-1}x\sqrt{-g}\Big[\phi R-\frac{\zeta}{4}\frac{g^{\mu\nu}\partial_{\mu}\phi\partial_{\nu}\phi}{\phi}\Big]}+S_\mathrm{M}\Big[\Psi,g_{\mu\nu}\Big],
\end{eqnarray}
where 
\begin{eqnarray}
S_\mathrm{M}\Big[\Psi,g_{\mu\nu}\Big]\equiv\frac{\zeta}{8\kappa_{D}^2}
\int{d^{D-1}x\sqrt{-g} \Big[\frac{\Phi\Psi_{,i}\Psi^{,i}}{\Psi^2}\Big]}
\end{eqnarray}


If we relax the static condition and we add a gauge field $A_{i}$ and using $D\to D+1$, $\sigma=\phi$, and $\gamma=(D-2)\phi$, we require the KK action. Our proposed model is a static analogous of the KK unified scenario of electromagnetism and gravity. The symmetries of the action in Eq.~(\ref{result1}) are 
presented as follows. 

\par

\noindent
{\bf  Diffeomorphism invariance}: 
The Bergmann-Wagoner formulation of theory which described by the action in Eq.~(\ref{result1}) is invariant under 
the $(D-1)$-dimensional diffeomorphism transformations: 
\begin{equation}
x^{i} \to x^{i}+\zeta(x^k)\,.
\end{equation}
Under this transformation, the metric tensor $h_{ij} ( \subseteq V_{D-1})$ 
behaves like a rank 2 tensor. 

\par

\noindent
{\bf Gauge transformations along the compactified coordinate}: 
In addition, this theory 
is invariant under the following gauge transformations 
\begin{eqnarray}
t \to t+\beta(x^k) \,.
\end{eqnarray}
Thanks to this symmetry, it always exists a wide local freedom to choice for the origin in system of coordinates along the compactified direction.
\par

Moreover, we investigate the two specific cases. 

\noindent
{\bf Case (i) General relativity (GR) with a massless scalar field}: 
If we set $\mu=-\lambda$, we find
\begin{equation}
S_{D-1}=\int{\sqrt{h}d^{D-1}x\Big(\frac{R_{D-1}}{2\kappa_D^2}+\frac{2(D-2)}{\kappa_D^2(D-3)}\mu_{,i}\mu^{,i}\Big)}-\frac{2\kappa_{D}^{-2}}{D-3}\int_{\partial M}{\sqrt{h}h^{ij}n_{i}\mu_{,j}d^{D-2}x}.
\end{equation}
It is just the GR in the presence of a massless scalar field with the right sign of the kinetic term.
\par
\noindent
 {\bf Case (ii) The KK theory in the Einstein frame without the $U(1)$ gauge field}: 
With $D\to D+1$, $\sigma=\phi$, $\gamma=(D-2)\phi$, and $A_{\mu}=0$, we get 
\begin{equation}
S_{D-1}=\int d^{D}x\sqrt{h}  \phi\left(\frac{R_{D}}{2\kappa_D^2}-\frac{3}{2}(D-1)(D-2)\partial_i\phi\partial^i\phi\right) \,.
\end{equation}
In this case, the scalar field $\phi$ is called the dilaton. 
To understand its meaning in a better way, we assume that there exists an $S^1$ in the vicinity of any point $x^\mu$ of $V_{D-1}$. The reduced metric on $S^1$ reads 
\begin{equation}
ds^2|_{S^1}=e^{2\gamma}dt^2 \,.
\end{equation}
The size of the circle is defined by 

\begin{equation}
\int_{S^1} \equiv \int_{0}^{2\pi R}ds_{S^1} = 2\pi R e^{\gamma}
\end{equation}

According to the terminology in the KK theory, we observe that the effective radius of $S^1$ at any point $x^{\mu}$ is $R e^{\gamma}$. 
Here, the dilaton field $\gamma$ defines the size of the $S^1$, 
similar to that in the KK theory. 
\par

\subsection{Equations of motion}

To find the equation of motion of the gravitational field 
for the model in Eq.~(\ref{result1}), we perform the variation 
of $S_{D-1}$ with respect to $h^{ij}$. 
The equations of motions are derived as 
\begin{eqnarray}
&&
r_{ij}-\frac{1}{2}rh_{ij} = \frac{1}{\Phi}\frac{\zeta}{8}\frac{\omega_\mathrm{BD}\Phi}{\Psi^2}\left( 2\Psi_{,i}\Psi_{,j}-h_{ij} \Psi_{,k}\Psi^{,k} 
\right) 
\nonumber \\ 
&& \hspace{25mm}
+\frac{\omega_\mathrm{BD}}{\Phi^2}\left(\Phi_{,i}\Phi_{,j}-\frac{1}{2}h_{ij}\Phi_{,k}\Phi^{,k}\right)+\frac{1}{\Phi}\left(\Phi_{,i,j}-h_{ij}\bigtriangleup\Phi\right)\,,
\label{eom1} \\
&&
\bigtriangleup\Phi = \frac{\kappa_n^2}{3+2\omega}\left[ \frac{\omega_\mathrm{BD}(3-D)}{2}\left(\frac{\zeta}{4}\frac{\partial_k\Psi\partial^k\Psi}{\Psi^2}+\frac{\partial_k\Phi\partial^k\Phi}{\Phi^2}\right)+(2-D)\frac{\bigtriangleup\Phi}{\Phi} \right]\,,
\label{eom2} \\
&&\frac{1}{\sqrt{h}}\partial_i\Big(\frac{\sqrt{h}\Phi h^{ij}\partial_j\Phi}{\Psi^2}\Big)+\frac{2\Phi}{\Psi^3}\partial_i\Psi\partial^i\Psi=0 \,.
\label{eom3}
\end{eqnarray}
Here, 
$r_{ij}=R_{ij} (h)$ 
corresponds to the Ricci tensor, which is constructed by $h_{ij}$, 
$r=R(h)=R(h)_{i}^{i}$, and $\omega_\mathrm{BD} = \frac{\zeta}{4}$.

\section{Successful reductions}

\subsection{General description}

Now, we suppose that the metric $g_{\mu\nu}$ is defined in the $D\to(D+p)$-dimensional space-time as
\begin{eqnarray}
V_{D+p} \Eqn{=} \cup_{i=1}^{p}V_{i}\oplus\cup_{j=p}^{D+p} V_j \,, 
\\
ds_{D+p}^2 \Eqn{=} g_{\mu\nu}dx^{\mu}\otimes dx^{\nu}=-\Sigma_{j=1}^{p} e^{2\alpha_j}dt_j\otimes dt_j+e^{2\Sigma_{j=1}^{p} \sigma_j}h_{AB}dx^A\otimes dx^B \,, 
\label{gpD} \\
\{A,B\} \Eqn{=} \{p,p+1,...,p+D\},\ 
\quad 
\alpha_j=2\left(\Sigma_{l=1}^{j}\gamma_l+\Sigma_{l=1}^{j-1}\sigma_l \right)\,.
\end{eqnarray}
We define the concept of $p$-static metric. We call that the metric $g_{\mu\nu}$ of the space-time is $p$-static, if there is a set of coordinates 
$x^{a}$ with $1\leq a\leq p$ such that 
\begin{equation}
g_{\mu,a}=0\,,
\quad 
\frac{\partial g_{\mu\nu}}{\partial x^a}=0\,. 
\label{p-static}
\end{equation}
Equivalently, there exists a set of the commutative Killing vectors 
$\zeta^{a}$ as 
\begin{equation}
\zeta^{a}\zeta^{b}-\zeta^{b}\zeta^{a}=0\,.
\end{equation}
These vector fields are generators of the symmetries of 
the metric $g_{\mu\nu}$.

\subsection{Concrete examples}

We examine the following two examples. 

\par
\noindent
{\bf Example (i) Flat Bianchi-I cosmological  models}: 
The Bianchi-I metric in four-dimensions is $p=3$-static in $D+p=4$ 
\begin{equation}
ds^2=-dt^2+\Sigma_{i=1}^{3}A_{i}(t)^2(dx^{i})^2 \,.
\end{equation}
The reason is, it is possible  to find a set of coordinates planar $x^a=\{x^1,x^2,x^3\}$ satisfying 
the $p$-static condition in Eq.~(\ref{p-static}). 
The Killing vectors correspond to these coordinates are 
\begin{equation}
\zeta^a=\{\partial_x,\partial_y,\partial_z\} \,.
\end{equation}
These vectors lead to the following algebraic structure:
\begin{equation}
[\zeta^x,\zeta^y]=[\zeta^y,\zeta^z]=[\zeta^z,\zeta^x]=0 \,.
\end{equation}
\par

\noindent
{\bf Example (ii) Static-spherically symmetric space-time}: 
The generally static-spherically symmetric metric in four-dimensions 
is $p=2$-static in $D+p=4$ 
\begin{equation}
ds^2=-A(r)dt^2+B(r)dr^2+C(r)(d\theta^2+\sin\theta^2 d\varphi^2)\,, 
\end{equation}
because we have a set of coordinates $x^a=\{t,\varphi\}$ for 
the $p$-static condition in Eq.~(\ref{p-static}). 
The set of commutative Killing vectors are 
\begin{equation}
\zeta^1=\partial_t,\ \ \zeta^{2}=\partial_{\varphi} \,.
\end{equation}
Using Eq.~(\ref{gpD}), we are able to reduce the action $S_{p+D}$ to the lower $D$ dimensional one. 
We assume that by Eq.~(\ref{gpD}), 
the metric $h_{AB}$ is static with respect to all the ``time'' 
coordinates $t_{j}$ as follows 
\begin{equation}
h_{A,t_j}=0\,, 
\quad 
\frac{\partial h_{AB}}{\partial t_j}=0\,, 
\quad
1\leq j\leq p \,.
\end{equation}
In the so-called string frame (the Jordan frame), 
when $\sigma_l = 0$ with $l=1, \dots, p$ the level of $p$-static, 
the Ricci scalar of $V_{D+p}$ and that of $V_{D}$ satisfy 
the equation 
\begin{equation}
R_{D+p}=R_{D}+2\Gamma_{,A}\Gamma^{,A}\,, 
\quad 
\Gamma_{,A}\Gamma^{,A}=\Sigma_{l=1}^{p}\partial_{,A}
\gamma_{l}\partial^{,A}\gamma_{l} \,.
\end{equation}
Hence, we obtain the following reduction from $S_{D+p}$ to $S_{D}$ 
\begin{equation}
S_{D}=\frac{S_{D+p}}{\int{\Pi_{j=1}^{p}dt_j}}=\int{\frac{e^{\Sigma_{j=1}^{p}\alpha_j}\sqrt{h}d^D x}{2\kappa^2_{D+p}}} \left(R_{D}+2\Gamma_{,A}\Gamma^{,A}
\right)\,,
\end{equation}
where in the second equality, the Lagrangian function in $S_{D}$ is independent of the coordinates $t_{j}$ with $1\leq j\leq p$. 
Another equivalent form is given by 
\begin{equation}
S_{D}=\int{\frac{e^{\Sigma_{j=1}^{p}\alpha_j}\sqrt{h}d^D x}{2\kappa^2_{D}}}\Big(R_{D}+2\Gamma_{,A}\Gamma^{,A}\Big),\ \ \kappa_{D}^2=\frac{\kappa_{D+p}^2}{\int \Pi_{j=1}^{p} dt_j}\,.
\end{equation}
If $\Sigma_{j=1}^{p}\alpha_j=0$, 
the model becomes a good and simple example. 
In this case, since the action of $S_{D}$ equals to the Einstein gravity with an auxiliary scalar field, all these fields live on the $D$-dimensional sub space-time
\begin{equation}
S_{D}=\int{\frac{\sqrt{h}d^D x}{2\kappa^2_{D}}}\Big(R_{D}+2\Gamma_{,A}\Gamma^{,A}\Big)\,.
\end{equation}
Thus, we conclude that if we start with a curved spec-time 
in $(D+p)$-dimensions, and if this space-time is $p$-static, after $p$-times reduction, by the suitable choices of the metric functions, we obtain the Einstein gravity with the scalar field in $D$-dimensions.

\subsection{Compactification mechanism}

For $p$-static metrics, provided that the following compactification has 
already been carried out
\begin{equation}
t_j\to t_j+\beta_j(x^k)\,, 
\quad 
1\leq j\leq p \,.
\end{equation}
This indicates that we have $p$ gauge freedoms to select the origin 
along the compactified directions for the simple reduction of 
$V_{D+p} \to V_{D}$.

\section{Perturbations of $AdS_D$}
\par

In this section, we study the first order linear stability of the $AdS_D$ in the framework of our reduced scalar-tensor model $\{h_{ij},\Psi,\Phi\}$.  Before we mention here that in our model, like BD, we can find black holes with cosmological constant (negative or positive) \cite{Gao:2006dx}. 

The description ``$AdS_D$'' denotes an exact solution, the so-called 
``the anti-de Sitter solution'', of the $D$-dimensional Einstein gravity with the negative cosmological constant. 
The end point of the instability of AdS is a Schwarzschild-AdS black hole, which is asymptotically stable. The (in)stability of AdS under the perturbations has recently been investigated from different points of the view~\cite{Bizon:2013gxa}. 
As we know, there are two regimes of perturbations: One is linear in which we study the time evolution of a scalar field in the AdS background. By solving the wave equation, we determine whether the AdS background is unstable. The second one is non-linear perturbations. Here, we perturb the fully back-reacted metric. The analysis of this case is more complex. By a numerical integration, we observe that \textit{``generic perturbations start to grow
rapidly after a certain time''}. We conclude that the AdS remains unstable.

\par 

Generally, there are three major types of perturbations for the AdS black hole, namely scalar, gravitational and electromagnetic perturbations. These perturbations are governed by a type of frequencies so called as quasi-normal modes (QNMs). They obtained by solving the Klein-Gordon equation and by an appropriate boundary condition imposed at horizon and infinity. It is believed that QNMs at the regime of ultra violet (UV) are proportional to the size of horizon. In $D$-dimensional AdS background, there is an analytical approach to find these QNMs, using the standard method of separation of the variables of wave equation. For $D\geq 4$ we need to use numerical methods. Indeed, 
these were often approximate solutions, and in different kinds of perturbations  (see~\cite{Papantonopoulos:2011zz} for a comprehensive review of methods). In the UV regime, or gravitational (tensor) and electromagnetic (vector) perturbations it is needed to consider the problem using perturbation's scheme. However, scalar perturbations for lower dimensions $D=3$ have analytic solutions. The most important modes are live in infra red (IR) regime, when the modes have the longest living time. The important result is that for $AdS_5$, and in the case of hyperbolic topology of horizon $\mathcal{H}^{D-2}/\Gamma$($\Gamma$ is a discrete group of isometries) scalar modes live longer than other types of perturbations. It has an important role in plasma. For tensor perturbations the case is completely different and more complex. Indeed, according to the AdS/CFT conjecture, since the harmonics are conserved and traceless on $AdS_D$, so there is no way to construct a perturbation theory to compute the IR frequencies. However, in our KK approach which we identify the AdS black hole as an exact solution for our scalar-tensor theory, there is a way to consider the fully back-reacted solutions as well. In our study we investigate the linear perturbations of AdS.\par

We describe the unperturbed (i.e., background) AdS solution by the following set of the functions 
\begin{equation}
\Phi=\Phi^{(0)}(y)\,, 
\quad 
\Psi=\Psi^{(0)}(y)\,, 
\quad 
h_{ij}^{(0)}=\delta_{ij}= \mathrm{diag} (1,1, \dots, 1)\,, 
\quad 
(n-1) \, \text{times}, 
\end{equation}
where $y (>0)$ is the radial coordinate of the half plane (one patch description of the AdS solution) and the metric is $(n-1)$-static. This set of functions satisfy equations of motion (\ref{eom1}-\ref{eom3}).  
The perturbations are expressed as 
\begin{eqnarray}
&&
\Psi^{(1)}=\Psi^{(0)}+\delta\Psi(y,\vec{x})\,,
\\
&&
\Phi^{(1)}=\Phi^{(0)}+\delta\Phi(y,\vec{x})\,,
\\
&&
h_{ij}^{(1)}=h_{ij}^{(0)}+\delta h_{ij}(y,\vec{x})\,,
\\
&&
\delta h_{ij}(y,\vec{x})=\mathrm{diag}\left[\delta h_{yy}(y,\vec{x}),\delta h_{11}(y,\vec{x}),..,\delta h_{(n-2)(n-2)}(y,\vec{x})\right]\,.
\end{eqnarray}
By computing the $\delta S_{D-1}=0$, we find the perturbations equations 
\begin{eqnarray}
&&
\frac{\zeta \Phi^{(0)}}{8}\left(\partial_i\ln\Psi^{(0)}\partial^i\ln \Psi^{(0)}
-\partial_i\ln\Phi^{(0)}\partial^i\ln \Phi^{(0)}\right)h^{(0)}_{ij}
\delta h^{ij}
\nonumber \\ 
&&
{}-\partial_{k}\Phi^{(0)}\left(h^{ij(0)}\delta\Gamma^{k}_{ji}
-h^{ik(0)}\delta \Gamma^{m}_{mi}\right)=0\,,
\label{deltah}\\
&&
\frac{1}{2}\left(\frac{\partial_i\Psi^{(0)}\partial^i\Psi^{(0)}}{(\Psi^{(0)})^2}-\frac{\partial_i\Phi^{(0)}\partial^i\Phi^{(0)}}{(\Phi^{(0)})^2}\right)\delta \Phi-\Phi^{(0)} \partial_i(\ln\Phi^{(0)})\delta(\partial^i(\ln\Phi))=0\,,
\label{deltaphi}\\
&&\partial_i(\ln\Psi^{(0)})\delta\left[\partial^i(\ln\Psi)\right]=0\,,
\label{deltapsi}
\end{eqnarray}
with 
\begin{equation}
\delta\Gamma_{ij}^k=\frac{1}{2}\delta^{kd}\left(\partial_j(\delta h_{id})+\partial_{i}(\delta h_{jd})-\partial_{d}(\delta h_{ij})\right)\,.
\end{equation}
Here, we have simplified the equations for the AdS solution 
due to the facts that for the AdS solution, we have $r_{ij}=0$, $r=0$, and 
$h=1$.
\par

Equations (\ref{deltaphi}) and (\ref{deltapsi}) for the AdS solutions 
give us the exact solutions in terms of the unperturbed  fields
\begin{eqnarray}
&&
\delta\Psi(y,\vec{x})=c_{0}(\vec{x})\Psi^{(0)}(y)\,, 
\\
&&
\delta\Phi(y,\vec{x})=c_1(\vec{x}) \exp\left[ \frac{1}{2}\int{ dy \frac{\frac{\partial_i\Psi^{(0)}\partial^i\Psi^{(0)}}{(\Psi^{(0)})^2}+\frac{\partial_i\Phi^{(0)}\partial^i\Phi^{(0)}}{(\Phi^{(0)})^2}}{\partial_y\ln\Phi^{(0)}}} \right]\,.
\end{eqnarray}
The form of Eq.~(\ref{deltah}) reads
\begin{equation}
\frac{\zeta}{8}\left[\frac{\frac{\partial_i\Psi^{(0)}\partial^i\Psi^{(0)}}{(\Psi^{(0)})^2}-\frac{\partial_i\Phi^{(0)}\partial^i\Phi^{(0)}}{(\Phi^{(0)})^2}}{\partial_y\ln\Phi^{(0)}}\right]\delta_{ij} \delta h^{ij}-\left[\partial_y(\delta h_{iy}-\delta h_{ii})+\partial_{\vec{x}}(\delta h_{\vec{x}y})\right]=0\,.
\label{delta,x}
\end{equation}
Thanks to the planar symmetry of unperturbed metric, we use the Fourier decomposition of the different components of the metric as follows: 
\begin{equation}
\delta h_{ij}(y,\vec{x})=
\int {d^{n-2}k e^{i\vec{k}.\vec{x}}\delta\tilde{h}_{ij}(y,\vec{k})}\,.
\end{equation}
Substituting this expression into Eq.~(\ref{delta,x}), we find 
\begin{equation}
g(y)\delta_{ij} \delta \tilde{h}^{ij}(y,\vec{k})-\left[\partial_y(\delta \tilde{h}_{iy}(y,\vec{k})-\delta \tilde{h}_{ii}(y,\vec{k}))+i\vec{k} \delta \tilde{h}_{\vec{x}y}(y,\vec{k}) \right]=0\,,
\end{equation}
where
\begin{equation}
g(y)=\frac{\zeta}{8}\left(\frac{\frac{\partial_i\Psi^{(0)}\partial^i\Psi^{(0)}}{(\Psi^{(0)})^2}-\frac{\partial_i\Phi^{(0)}\partial^i\Phi^{(0)}}{(\Phi^{(0)})^2}}{\partial_y\ln\Phi^{(0)}}\right)\,.
\end{equation}
With diagonal metric $\delta h_{ij}(y,\vec{k})$, we obtain
\begin{equation}
g(y)\delta_{ij}  \delta \tilde{h}^{ij}(y,\vec{k})+\partial_y(\delta \tilde{h}_{ii}(y,\vec{k}))=0\,.
\end{equation}
A simple integration leads to 
\begin{equation}
\delta_{ij}  \delta \tilde{h}^{ij}(y,\vec{k})=
c_{2}(\vec{k})e^{-\int{g(y)dy}}\,.
\end{equation}
Consequently, we get
\begin{equation}
\delta h_{ij}(y,\vec{x})=e^{-\int{g(y)dy}}\int {d^{D-2}k e^{i\vec{k}.\vec{x}}c_2(\vec{k})}=e^{-\int{g(y)dy}}\tilde{c}_{ij}(\vec{x})\,.
\end{equation}
To complete the stability analysis of the AdS solution, 
by making use of the relations 
$\Psi^{(0)}=y^{-1}$, $\Phi^{(0)}=y^{2-D}$, and 
$g(y)=\frac{\zeta(-D^2+4D-3)}{8(2-D)}y^{-1}$, we solve equations of motion 
(\ref{eom1})--(\ref{eom3}) softly. 
As a result, we acquire
\begin{equation}
\delta h_{ij}(y,\vec{x})=y^{\frac{8(2-D)}{\zeta(-D^2+4D-3)}}
\tilde{c}_{ij}(\vec{x})\,.
\end{equation}
In order for the AdS solution to be stable, 
the perturbations have to tend to zero at the asymptotic boundary $y\to 0$, where $|\tilde{c}_{ij}(\vec{x})| \to 0$. 
Thus, the following conditions are required
\begin{equation}
D>1 \,.
\end{equation}
However, since in our reduction, we have $D>5$, for this range we find 
\begin{eqnarray}
&&\lim _{y\to 0} \delta h_{ij}(y,\vec{x})=\tilde{c}_{ij}(\vec{x})\lim _{y\to 0}y^{\frac{4(D-3)}{(D^2-4D+3)}}=0\,, \\
&&
\lim _{y\to 0}\delta\Psi(y,\vec{x})=c_{0}(\vec{x})y^{-1}=\infty\,, \\
&&
\lim _{y\to 0}\delta\Phi(y,\vec{x})=c_1(\vec{x}) y^{\frac{1}{2}\frac{D^2-4D+5}{2-D}}=\infty\,.
\end{eqnarray}

Accordingly, 
we conclude that $AdS_D$ is unstable in our scalar model. 
This result is supported by different methods in Ref.~\cite{Bizon:2013gxa}. 
Indeed, $AdS_{D+1}$, where $D \geq 3$, is unstable against a large class of 
arbitrarily small perturbations. 
Our stability analysis has been performed for one patch of the AdS space. 
To complete the analysis of the AdS solution, 
we suppose that the unperturbed $AdS_D$ is parametrized 
by the following set of the functions 
\begin{eqnarray}
&&
\Phi^{(0)}(x) = \Big(\frac{l}{\cos x}\Big)^{D-2}\,,\\
&&
\Psi^{(0)}(x) = (\frac{l}{\cos x}\Big)\,,\\
&&
h_{ij}^{(0)} = \mathrm{diag} (1,\sin^2 x\omega_{D-2})\,, 
\end{eqnarray}
where $(t,x)\in \mathcal{R}\times[0,\pi/2)$, and $\omega_{D-2}$ is the metric of a unit sphere $S^{D-2}$. 
The perturbations are expressed as 
\begin{eqnarray}
&&
\Psi^{(1)}=\Psi^{(0)}+\delta\Psi(t,x,\omega)\,, 
\label{eq:3.25}\\
&&
\Phi^{(1)}=\Phi^{(0)}+\delta\Phi(t,x,\omega)\,,
\label{eq:3.26}\\
&&
h_{ij}^{(1)}=h_{ij}^{(0)}+\delta h_{ij}(t,x,\omega)\,,
\label{eq:3.27}\\
&&
\delta h_{ij}(t,x,\omega)=\mathrm{diag}\left[\delta h_{xx}(t,x,\omega),\delta h_{11}(t,x,\omega),..,\delta h_{(D-2)(D-2)}(t,x,\omega)\right]\,.
\end{eqnarray}
We represent all the perturbations in terms of the harmonics $Y_k(\omega)$ 
on the $S^{D-2}$ as 
\begin{equation}
\Delta_{\omega}Y_k(\omega)=-l(l+D-3)Y_k(\omega)\,.
\end{equation}
Hence, we rewrite Eqs.~(\ref{eq:3.25})--(\ref{eq:3.27}) to 
\begin{eqnarray}
&&
\Psi^{(1)}=\Psi^{(0)}+\Sigma_k\delta\Psi_k(t,x)Y_k(\omega)\,,
\\
&&
\Phi^{(1)}=\Phi^{(0)}+\Sigma_k\delta\Phi_k(t,x)Y_k(\omega)\,,
\\
&&
h_{ij}^{(1)}=h_{ij}^{(0)}+\Sigma_k\delta h^{k}_{ij}(t,x)Y_k(\omega)\,.
\end{eqnarray}
The solution of perturbations is given by 
\begin{equation}
\delta\Psi_k(t,x)=c^{+}_k(t)\frac{e^{\frac{l \sin x}{2\cos^2 x}}}{(\frac{1+\sin x}{\cos x})^{l/2}}\,.
\label{eq:3.33}
\end{equation}
This result is valid for the electromagnetic and gravitational perturbations. 
If we define $y \equiv \pi/2-x$, Eq.~(\ref{eq:3.33}) is described as 
\begin{equation}
\delta\Psi_k(t,y)=c^{+}_k(t)\frac{e^{\frac{l \cos y}{2\sin^2 y}}}{(\frac{1+\cos y}{\sin y})^{l/2}}\,.
\end{equation}
We need to obtain $\lim_{y \to 0} \delta\Psi_k(t,y) \to 0$. 
In fact, we acquire 
\begin{equation}
\lim_{y \to 0} \delta\Psi_k(t,y)=c^{+}_k(t) 
\exp \left[ \frac{l}{2y^2}+\frac{l}{6}+\frac{l}{2} \left(\ln y-\ln 2 
-\frac{1}{2}\right)+\mathcal{O}(y^2) \right] \to \infty \,.
\end{equation}

Furthermore, we get
\begin{eqnarray}
\delta\Phi_k(t,x) \Eqn{=} c^{-}_k(t) \exp\left[ \frac{1}{2}\int{ dx \frac{\frac{\partial_i\Psi^{(0)}\partial^i\Psi^{(0)}}{(\Psi^{(0)})^2}+\frac{\partial_i\Phi^{(0)}\partial^i\Phi^{(0)}}{(\Phi^{(0)})^2}}{\partial_x\ln\Phi^{(0)}}} \right] 
\nonumber \\
\Eqn{=} c^{-}_k(t) \Big(\cos x\Big)^{-\frac{1}{2}\frac{D^2-3D+4}{D-2}}=c^{-}_k(t) \Big(\sin y\Big)^{-\frac{1}{2}\frac{D^2-3D+4}{D-2}}\,.
\end{eqnarray}
The only possibility to have $\lim_{y \to 0} \delta\Phi_k(t,y)=0$ 
is $D<2$, the condition of which is physically unacceptable. 
It is possible to integrate and find $\delta h^k_{ij}(t,x)$. 
Thus, the AdS solution is unstable under the non-linear perturbations 
in terms of $\Psi$. 
This consequence proves that in a similar sense, 
the former results on the instability of the $AdS_D$ under the non-linear perturbations are obtained.

\par

\section{Conclusions}

In the present paper, 
we have presented the formulation of the KK dimensional reduction from higher-dimensional space-time. Particularly, we have explored the resultant Bergmann-Wagoner general action of  scalar-tensor theories. 
Furthermore, we have derived the conditions that 
the dimensional reduction can successfully be realized. The consequence of the KK compactification has also been stated.  

As an instructive example we have investigated the stability of the AdS solution by examining the perturbations from the background AdS solution. 
In our model, the AdS space-time exists as an exact solution to the field equations given by (\ref{eom1})--(\ref{eom3}), and it is unstable 
under the perturbations. 
We compared our results with the previously results of QNMs, we showed that since the most important modes are live in infrared (IR) regime, with the longest living time, in the case of hyperbolic topology of horizon which we studied in our paper, scalar modes live longer than other types of perturbations. We have investigate the fully back-reacted AdS background using our scalar-tensor picture. As we have been shown that explicitly, the analysis of tensor perturbations in this case is completely different from the other modes. However, in our KK approach which we identify the AdS black hole as an exact solution for the Bergmann-Wagoner scalar-tensor theory, there is a way to consider the fully back-reacted solutions as well. We have been proven that AdS solution still remains unstable. This result is valid for both patches of the AdS.

\section*{Acknowledgments}

The work has been supported in part by the JSPS Grant-in-Aid
for Young Scientists (B) \# 25800136 and the research-funds presented 
by Fukushima University (K.B.).


\end{document}